\documentclass[10pt]{iopart}

\usepackage{graphicx}% Include figure files
\usepackage{dcolumn}% Align table columns on decimal point
\usepackage{bm}% bold math
\usepackage{color}

\begin{document}

\title[Alfven Eigenmodes in ITER]{Analysis of Alfven Eigenmode destabilization in ITER using a Landau closure model}

%\address{IOP Publishing, Temple Circus, Temple Way, Bristol BS1 6HG, UK}
%\ead{submissions@iop.org}

\author{J. Varela}
\ead{jacobo.varela@nifs.ac.jp}
\address{National Institute for Fusion Science, National Institute of Natural Science, Toki, 509-5292, Japan}
\address{Oak Ridge National Laboratory, Oak Ridge, Tennessee 37831-8071, USA}
\author{D. A. Spong}
\address{Oak Ridge National Laboratory, Oak Ridge, Tennessee 37831-8071, USA}
\author{L. Garcia}
\address{Universidad Carlos III de Madrid, 28911 Leganes, Madrid, Spain}

\date{\today}

\begin{abstract}
Alfv\' en Eigenmodes (AE) can be destabilized during ITER discharges driven by neutral beam injection (NBI) energetic particles (EP) and alpha particles. The aim of the present study is to analyze the AE stability of different ITER operation scenarios considering multiple energetic particle species. We use the reduced magneto-hydrodynamic (MHD) equations to describe the linear evolution of the poloidal flux and the toroidal component of the vorticity in a full 3D system, coupled with equations of density and parallel velocity moments for the energetic particles species including the effect of the acoustic modes. The AEs driven by the NBI EP and alpha particles are stable in the configurations analyzed, only MHD-like modes with large toroidal couplings are unstable, although both can be destabilized if the EP $\beta$ increases above a threshold. The threshold is two times the model $\beta$ value for the NBI EP and alpha particles in the reverse shear case, leading to the destabilization of Beta induced AE (BAE) near the magnetic axis with a frequency of $25-35$ kHz and Toroidal or Elliptical AE (TAE/EAE) in the reverse shear region with a frequency of $125-175$ kHz, respectively. On the other hand, the hybrid and steady state configurations show a threshold $3$ times larger with respect to the model $\beta$ for the alpha particle and $40$ times for the NBI EP, also destabilizing BAE and TAE between the inner and middle plasma region. In addition, a extended analysis of the reverse shear scenario where the $\beta$ of both alpha particles and NBI EP are above the AE threshold, multiple EP damping effects are also identified as well as optimization trends regarding the resonance properties of the alpha particle and NBI EP with the bulk plasma.

\end{abstract}

%\maketitle

\pacs{52.35.Py, 52.55.Hc, 52.55.Tn, 52.65.Kj}% PACS, the Physics and Astronomy

\vspace{2pc}
\noindent{\it Keywords}: Tokamak, DIII-D, Pedestal, MHD, AE, energetic particles

This manuscript has been authored by UT-Battelle, LLC under Contract No. DE-AC05- 00OR22725 with the U.S. Department of Energy. The United States Government retains and the publisher, by accepting the article for publication, acknowledges that the United States Government retains a non-exclusive, paid-up, irrevocable, world-wide license to publish or reproduce the published form of this manuscript, or allow others to do so, for United States Government purposes. The Department of Energy will provide public access to these results of federally sponsored research in accordance with the DOE Public Access Plan (http://energy.gov/downloads/doe-public-access-plan).

\maketitle

\ioptwocol

\section{Introduction \label{sec:introduction}}

Energetic particle (EP) driven instabilities can enhance the transport of fusion produced alpha particles, energetic neutral beams and particles heated using ion cyclotron resonance heating (ICRF) \cite{1,2,3}. The consequence is a decrease of the heating efficiency in devices  as TFTR, JET and DIII-D tokamaks or LHD, TJ-II and W7-AS stellarators \cite{4,5,6,7,8,9}. If the mode frequency resonates with the drift, bounce or transit frequencies of the energetic particles, the particle and diffusive losses increase. In addition, plasma instabilities such as internal kinks \cite{10,11} or ballooning modes \cite{12} can be kinetically destabilized.

Alfv\' en Eigenmodes (AE) are driven in the spectral gaps of the shear Alfv\' en continua \cite{13,14}, destabilized by super-Alfv\' enic alpha particles and energetic particles. Alfv\' en Eigenmode (AE) activity was observed before in several discharges and configurations \cite{15,16,17,18}. The different Alfv\' en eigenmode families ($n$ is the toroidal mode and $m$ the poloidal mode) are linked to frequency gaps produced by periodic variations of the Alfv\' en speed, for example: toroidicity induced Alfv\' en Eigenmodes (TAE) couple $m$ with $m+1$ modes \cite{19,20,21}, beta induced Alfv\' en Eigenmodes driven by compressibility effects (BAE) \cite{22}, Reversed-shear Alfv\' en Eigenmodes (RSAE) due to local maxima/minima in the safety factor $q$ profile \cite{23}, Global Alfv\' en Eigenmodes (GAE) observed in the minimum of the Alfv\' en continua \cite{24,25}, ellipticity induced Alfv\' en Eigenmodes (EAE) coupling $m$ with $m+2$ modes \cite{26,27} and noncircularity induced Alfv\' en Eigenmodes (NAE) coupling $m$ with $m+3$ or higher \cite{28,29}.

The ITER plasma will be heated by two NBI's delivering in the plasma $33.3$ MW of 1 MeV $D^{0}$ or $0.87$ MeV $H^{0}$ \cite{30}. Several theoretical studies and extrapolations from other devices observations suggest the destabilization of BAE, TAE and EAE during the ITER discharges \cite{31,32,33,34,35,36,37}. It should be noted that different EP species will coexist in the plasma as the NBI driven EP and alpha particles. The destabilizing effect of the combined EP species populations has not been extensively studied, although it is a critical issue in ITER \cite{38,39,40,41}, so it is desirable to analyze the AE stability of a plasma with multiple EP species. Experiments in the TFTR device already identified the multiple EP species effect in the plasma AE stability; alpha particle driven AEs were stabilized by the presence of the NBI driven EP species, only measured at the end of the discharge after the beam injection was stopped \cite{42,43,44}. 

Several advanced operation scenarios have been proposed for ITER \cite{45,46,47,48,49,50,51}. Among them, the reverse shear \cite{52,53,54}, hybrid \cite{55,56,57} and steady state operation \cite{58,59,60} scenarios are the configurations with the best potential performance. The hybrid scenario was already tested in several tokamaks showing an improved confinement due to the low level of MHD instabilities, associated with a flat safety factor profile in the plasma core \cite{61,62,63}. On the other hand, reverse shear configurations were proposed as a possible steady state operation scenario \cite{64,65,66,67}, because such configurations can sustain a large fraction of aligned bootstrap current \cite{68,69,70} and improved MHD stability \cite{71,72,73}.

The aim of the present study is to analyze the AE stability of ITER plasma, identifying the configurations with damping on the dominant AE caused by multiple EP species effects. If the AEs growth rate of the multiple EP species case (NBI EP + alpha particles) is smaller compared to the AEs destabilized individually by the NBI driven EP and alpha particles, the device is considered to be operated in the multiple EP interaction regime. In addition, we study the effect of the NBI EP and alpha particles $\beta$, energy and density profile on the instabilities growth rate and frequency.

A set of simulations are performed using the FAR3d code \cite{74,75,76}. The model includes the reduced MHD equations and the moment equations of the NBI driven energetic ion and alpha particles density and parallel velocity \cite{77,78}. The FAR3d code, with the appropriate Landau closure relations, solves the reduced linear resistive MHD equations including the linear wave-particle resonance effects, required for Landau damping/growth, and the parallel momentum response of the thermal plasma, required for coupling to the geodesic acoustic waves \cite{79}. The code follows the evolution of eight field variables, starting from an equilibria calculated by the VMEC code \cite{80}. A methodology has been developed to calibrate Landau-closure models against more complete kinetic models and optimize the closure coefficients \cite{79}. The model includes Landau resonance couplings, thermal ion and energetic particle FLR effects \cite{78} as well as the Landau damping of the modes on the background ions/electrons \cite{77}.

This paper is organized as follows. The model equations, numerical scheme and equilibrium properties are described in section \ref{sec:model}. The effect of the alpha particle $\beta$ on the instability properties is analyzed in section \ref{sec:beta}. The effect of the NBI EP $\beta$ on the instability properties is studied in section \ref{sec:beta2}. We analyze the instability thresholds if both NBI EP and alpha particles driven AE are unstable in section \ref{sec:threshold}. Finally, the conclusions of this paper are presented in section \ref{sec:conclusions}.

\section{Equations and numerical scheme \label{sec:model}}

A reduced set of equations to describe the evolution of the background plasma and fields, retaining the toroidal angle variation, are used in the present study. These are derived from the method employed in Ref.\cite{81} assuming high-aspect ratio configurations with moderate $\beta$-values. We obtain a reduced set of equations using the exact two (tokamak) or three-dimensional (stellarator) equilibrium. The effect of the energetic particle population is included in the formulation as moments of the kinetic equation truncated with a closure relation \cite{82}. These describe the evolution of the density ($n_{f}$) and velocity moments parallel to the magnetic field lines ($v_{||f}$) of the NBI driven EP and alpha particles. The coefficients of the closure relation are selected to match a two-pole approximation of the plasma dispersion function.    

The plasma velocity and perturbation of the magnetic field are defined as
\begin{equation}
 \mathbf{v} = \sqrt{g} R_0 \nabla \zeta \times \nabla \tilde\Phi, \quad\quad\quad  \mathbf{B} = R_0 \nabla \zeta \times \nabla \tilde\psi,
\end{equation}
where $\zeta$ is the toroidal angle, $\tilde\Phi$ is a perturbation of the stream function proportional to the electrostatic potential, and $\tilde\psi$ is the perturbation of the poloidal flux.

The equations, in dimensionless form, are
\begin{equation}
\label{poloidal flux}
\frac{\partial \tilde \psi}{\partial t} =  \sqrt{g} B \nabla_{||} \tilde\Phi  + \eta \varepsilon^2 J \tilde J^\zeta + \frac{S}{\epsilon^{2}} \frac{B_{0} \rho_{i}^{2}}{J-\rlap{-} \iota I} \frac{v_{A0}^{2}}{v_{Te}} \sqrt{\frac{\pi}{2}} | \nabla_{||}Q |
\end{equation}
\begin{eqnarray} 
\label{vorticity}
\frac{{\partial \tilde U}}{{\partial t}} =  -\epsilon v_{\zeta,eq} \frac{\partial U}{\partial \zeta} \nonumber\\
+ S^2 \left[{ \sqrt{g} B \nabla_{||} J^\zeta - \frac{\beta_0}{2\varepsilon^2} \sqrt{g} \left( \nabla \sqrt{g} \times \nabla \tilde p \right)^\zeta }\right]   \nonumber\\
-  S^2 \left[{\frac{\beta^{f,\alpha}}{2\varepsilon^2} \sqrt{g} \left( \nabla \sqrt{g} \times \nabla \tilde n^{f,\alpha} \right)^\zeta }\right] + S\omega_{r} \rho_{i}^{2} \nabla_{\perp}^{2} U
\end{eqnarray} 
\begin{eqnarray}
\label{pressure}
\frac{\partial \tilde p}{\partial t} = -\epsilon v_{\zeta,eq} \frac{\partial p}{\partial \zeta} + \frac{dp_{eq}}{d\rho}\frac{1}{\rho}\frac{\partial \tilde \Phi}{\partial \theta} \nonumber\\
 +  \Gamma p_{eq}  \left[{ \sqrt{g} \left( \nabla \sqrt{g} \times \nabla \tilde \Phi \right)^\zeta - \nabla_{||}  v_{|| th} }\right]
\end{eqnarray} 
\begin{eqnarray}
\label{velthermal}
\frac{{\partial \tilde v_{|| th}}}{{\partial t}} = -\epsilon v_{\zeta,eq} \frac{\partial v_{||th}}{\partial \zeta} -  \frac{S^2 \beta_0}{n_{0,th}} \nabla_{||} p 
\end{eqnarray}
\begin{eqnarray}
\label{nfast}
\frac{{\partial \tilde n}}{{\partial t}}^{f,\alpha} = -\epsilon v_{\zeta,eq} \frac{\partial n^{f,\alpha}}{\partial \zeta} - \frac{S  (v_{th}^{f,\alpha})^2}{\omega_{cy}^{f,\alpha}} \Omega_d (\tilde n^{f,\alpha}) \nonumber\\
- S  n_{0}^{f,\alpha} \nabla_{||} v_{||}^{f,\alpha} - \varepsilon^2  n_{0}^{f,\alpha} \Omega_d (\tilde \Phi) + \varepsilon^2 n_{0}^{f,\alpha} \Omega_{*} (\tilde  \Phi) \nonumber\\
+ \left[ \frac{\epsilon^{2}\omega_{cy}^{f,\alpha}}{(v_{th}^{f,\alpha})^2} \omega_{r} n_{0}^{f,\alpha} - \frac{1}{\rho (J-\rlap{-} \iota I)} \frac{dn_{0}^{f,\alpha}}{d\rho} \left(I \frac{\partial}{\partial \zeta} - J \frac{\partial}{\partial \theta} \right) \right] W^{f,\alpha}
\end{eqnarray}
\begin{eqnarray}
\label{vfast}
\frac{{\partial \tilde v_{||}}}{{\partial t}}^{f,\alpha} = -\epsilon v_{\zeta,eq} \frac{\partial v_{||}^{f,\alpha}}{\partial \zeta}  -  \frac{S  (v_{th}^{f,\alpha})^2}{\omega_{cy}^{f,\alpha}} \, \Omega_d (\tilde v_{||}^{f,\alpha}) \nonumber\\
- \left( \frac{\pi}{2} \right)^{1/2} S  v_{th}^{f,\alpha} \left| \nabla_{||} \right|  v_{||}^{f,\alpha} - \frac{S  (v_{th}^{f,\alpha})^2}{n_{0}^{f,\alpha}} \nabla_\| n^{f,\alpha}  \nonumber\\
+ \frac{S^2 (v_{th}^{f,\alpha})^2}{\rho (J-\rlap{-} \iota I)} \frac{1}{n_{0}^{f,\alpha}} \frac{dn_{0}^{f,\alpha}}{d\rho} \left(I X_{1}^{f,\alpha} - J X_{2}^{f,\alpha} \right)
\end{eqnarray}
Here, $U =  \sqrt g \left[{ \nabla  \times \left( {\rho _m \sqrt g {\bf{v}}} \right) }\right]^\zeta$ is the vorticity and $\rho_m$ the ion and electron mass density. The auxiliary equations are:
\begin{equation}
0 = Q - \nabla_{\perp}^{2}\tilde\psi
\end{equation}
\begin{equation}
0 = (1 - (\rho_{f,\alpha})^{2} \nabla_{\perp}^{2}) W^{f,\alpha} - (\rho_{f,\alpha})^{2} \nabla_{\perp}^{2} \tilde\Phi
\end{equation}
 \begin{equation}
0 = (1 - (\rho_{f,\alpha})^{2} \nabla_{\perp}^{2}) X_{1}^{f,\alpha} - \frac{\partial \tilde\psi}{\partial \zeta}
\end{equation}
 \begin{equation}
0 = (1 - (\rho_{f,\alpha})^{2} \nabla_{\perp}^{2}) X_{2}^{f,\alpha} - \frac{\partial \tilde\psi}{\partial \theta}
\end{equation}
with $\rho_{i}$ the finite Larmor radius of the thermal ions, $\rho_{f}$ the finite Larmor radius of the NBI driven EP and $\rho_{\alpha}$ the finite Larmor radius of the alpha particles (all normalized to the minor radius). The toroidal current density $J^{\zeta}$ is defined as:
\begin{eqnarray}
J^{\zeta} =  \frac{1}{\rho}\frac{\partial}{\partial \rho} \left(-\frac{g_{\rho\theta}}{\sqrt{g}}\frac{\partial \tilde\psi}{\partial \theta} + \rho \frac{g_{\theta\theta}}{\sqrt{g}}\frac{\partial \tilde\psi}{\partial \rho} \right) \nonumber\\
- \frac{1}{\rho} \frac{\partial}{\partial \theta} \left( \frac{g_{\rho\rho}}{\sqrt{g}}\frac{1}{\rho}\frac{\partial \tilde\psi}{\partial \theta} + \rho \frac{g_{\rho \theta}}{\sqrt{g}}\frac{\partial \tilde\psi}{\partial \rho} \right)
\end{eqnarray}
$v_{||th}$ is the parallel velocity of the thermal particles, $v_{\zeta,eq}$ is the equilibrium toroidal rotation and $v_{\perp} = - \nabla \tilde\Phi \times B^{\zeta} e_{\zeta} / B^2$ is the thermal perpendicular velocity. $n_{f}$ is normalized to the density at the magnetic axis $n_{f_{0}}$, $\tilde\Phi$ to $a^2B_{0}/\tau_{R}$ and $\tilde\psi$ to $a^2B_{0}$. All lengths are normalized to a generalized minor radius $a$; the resistivity to $\eta_0$ (its value at the magnetic axis); the time to the resistive time $\tau_R = a^2 \mu_0 / \eta_0$; the magnetic field to $B_0$ (the averaged value at the magnetic axis); and the pressure to its equilibrium value at the magnetic axis. The Lundquist number $S$ is the ratio of the resistive time to the Alfv\' en time $\tau_{A0} = R_0 (\mu_0 \rho_m)^{1/2} / B_0$. $\rlap{-} \iota$ is the rotational transform, $v_{th,f} = \sqrt{T_{f}/m_{f}}/v_{A0}$ the energetic particle thermal velocity normalized to the Alfv\' en velocity in the magnetic axis and $\omega_{cy}$ the energetic particle cyclotron frequency times $\tau_{A0}$. $q_{f}$ is the charge, $T_{f}$ the temperature and $m_{f}$ the mass of the energetic particles. The $\Omega$ operators are defined as:
\begin{eqnarray}
\label{eq:omedrift}
\Omega_d = \frac{1}{2 B^4 \sqrt{g}}  \left[  \left( \frac{I}{\rho} \frac{\partial B^2}{\partial \zeta} - J \frac{1}{\rho} \frac{\partial B^2}{\partial \theta} \right) \frac{\partial}{\partial \rho}\right] \nonumber\\
-   \frac{1}{2 B^4 \sqrt{g}} \left[ \left( \rho \beta_* \frac{\partial B^2}{\partial \zeta} - J \frac{\partial B^2}{\partial \rho} \right) \frac{1}{\rho} \frac{\partial}{\partial \theta} \right] \nonumber\\ 
+ \frac{1}{2 B^4 \sqrt{g}} \left[ \left( \rho \beta_* \frac{1}{\rho} \frac{\partial B^2}{\partial \theta} -  \frac{I}{\rho} \frac{\partial B^2}{\partial \rho} \right) \frac{\partial}{\partial \zeta} \right]
\end{eqnarray}
\begin{eqnarray}
\label{eq:omestar}
\Omega_* = \frac{1}{B^2 \sqrt{g}} \frac{1}{n_{f0}} \frac{d n_{f0}}{d \rho} \left( \frac{I}{\rho} \frac{\partial}{\partial \zeta} - J \frac{1}{\rho} \frac{\partial}{\partial \theta} \right) 
\end{eqnarray}
Here the $\Omega_{d}$ operator is constructed to model the average drift velocity of a passing particle and $\Omega_{*}$ models its diamagnetic drift frequency. We also define the parallel gradient and curvature operators:
\begin{equation}
\label{eq:gradpar}
\nabla_\| f = \frac{1}{B \sqrt{g}} \left( \frac{\partial \tilde f}{\partial \zeta} -  \rlap{-} \iota \frac{\partial \tilde f}{\partial \theta} - \frac{\partial f_{eq}}{\partial \rho}  \frac{1}{\rho} \frac{\partial \tilde \psi}{\partial \theta} + \frac{1}{\rho} \frac{\partial f_{eq}}{\partial \theta} \frac{\partial \tilde \psi}{\partial \rho} \right)
\end{equation}
\begin{equation}
\label{eq:curv}
\sqrt{g} \left( \nabla \sqrt{g} \times \nabla \tilde f \right)^\zeta = \frac{\partial \sqrt{g} }{\partial \rho}  \frac{1}{\rho} \frac{\partial \tilde f}{\partial \theta} - \frac{1}{\rho} \frac{\partial \sqrt{g} }{\partial \theta} \frac{\partial \tilde f}{\partial \rho}
\end{equation}
with the Jacobian of the transformation:
\begin{equation}
\label{eq:Jac}
\frac{1}{\sqrt{g}} = \frac{B^2}{\varepsilon^2 (J- \rlap{-} \iota I)}
\end{equation}

Equations~\ref{pressure} and~\ref{velthermal} introduce the parallel momentum response of the thermal plasma, required for coupling to the geodesic acoustic waves, accounting the geodesic compressibility in the frequency range of the geodesic acoustic mode (GAM) \cite{83,84}. The index $f$ and $\alpha$ indicates that the model includes individual equations for the density and parallel velocity of the NBI driven EP and alpha particles.

Equilibrium flux coordinates $(\rho, \theta, \zeta)$ are used. Here, $\rho$ is a generalized radial coordinate proportional to the square root of the toroidal flux function, and normalized to one at the edge. The flux coordinates used in the code are those described by Boozer \cite{85}, and $\sqrt g$ is the Jacobian of the coordinate transformation. All functions have equilibrium and perturbation components represented as: $ A = A_{eq} + \tilde{A} $. 

The FAR3d code uses finite differences in the radial direction and Fourier expansions in the two angular variables. The numerical scheme is semi-implicit in the linear terms. The nonlinear version uses a two semi-step method to ensure $(\Delta t)^2$ accuracy.

The present model was already used to study the AE activity in LHD \cite{86,87}, TJ-II \cite{88,89,90} and DIII-D \cite{91} indicating reasonable agreement with the observations.

\subsection{Equilibrium properties}

We use fixed boundary results from the VMEC equilibrium code \cite{80} calculated using TRANSP simulations. Due to the fact that the FAR3d stability model is based on stellarator symmetry, we null out the up-down asymmetric terms in the VMEC shape and base the calculations of the current paper on up-down symmetric equilibria. We analyze three operation scenarios of ITER: reverse shear (RS), hybrid (H) and steady state configurations (SS). Table~\ref{Table:1} shows the main parameters of the thermal plasma, table~\ref{Table:2} the NBI driven EP configurations and table~\ref{Table:3} the properties of the alpha particles for each operation scenario at the magnetic axis. The profiles used in this study are obtained from previous numerical analysis dedicated to foreseen future operation scenarios of ITER \cite{92,93,94,95}.

\begin{table}[t]
\centering
\begin{tabular}{c | c c c c}
Case & $T_{i}(0)$ & $n_{i}(0)$ & $\beta_{th}(0)$ & $V_{A0}$ \\
 & (keV) & ($10^{20}$ m$^{-3}$) &  & ($10^{7}$ m/s) \\ \hline
RS & 20 & 0.25 & 0.020 & 1.1 \\
H & 25 & 1.05 & 0.078 & 0.71 \\
SS & 36 & 0.7 & 0.080 & 0.87 \\
\end{tabular}
\caption{Thermal plasma properties (values at the magnetic axis). Column 1 is the thermal ion temperature, column 2 is the thermal ion density, column 3 is the thermal $\beta$ and column 4 is the Alfv\' en velocity.} \label{Table:1}
\end{table}

\begin{table}[t]
\centering
\begin{tabular}{c | c c c c}
Case & $n_{f}(0)$ & $T_{f}(0)$ & $\beta_{f}(0)$ & $\omega_{cy,f}$ \\
 &  ($10^{20}$ m$^{-3}$) & (keV) &  & \\ \hline
RS & 0.0023 & 200 & 0.00073 & 133 \\
H & 0.0023 & 200 & 0.00065 & 220 \\
SS & 0.0023 & 200 & 0.00065 & 180 \\
\end{tabular}
\caption{NBI driven energetic particles properties (values at the magnetic axis). Column 1 is the temperature, column 2 the density, column 3 the $\beta$ and column 4 the normalized cyclotron frequency (to the Alfv\' en time).} \label{Table:2}
\end{table}

\begin{table}[t]
\centering
\begin{tabular}{c | c c c c}
Case & $n_{\alpha}(0)$ & $T_{\alpha}(0)$ & $\beta_{\alpha}(0)$ & $\omega_{cy,\alpha}$ \\
 & ($10^{20}$ m$^{-3}$) & (keV) &  & \\ \hline
RS & 0.004 & 990 & 0.0062 & 66.32\\
H & 0.037 & 846 & 0.0094 & 110\\
SS & 0.41 & 997 & 0.0102 & 90\\
\end{tabular}
\caption{Alpha particles properties (values at the magnetic axis). Column 1 is the temperature, column 2 the density, column 3 the $\beta$ and column 4 the normalized cyclotron frequency (to the Alfv\' en time).} \label{Table:3}
\end{table}

For all the models, the $\beta_{0}$ of the NBI driven EP and alpha particles required to destabilize the AEs are above the values calculated using TRANSP (see table~\ref{Table:2} and \ref{Table:3}), indicating that the EP drive in the simulations is larger respect to the reference models.

The magnetic field at the magnetic axis is $5$ ($5.3$) T and the averaged inverse aspect ratio is $\varepsilon=0.41$ ($0.32$) in the RS (H and SS) case. 

\begin{figure*}[h!]
\centering
\includegraphics[width=0.8\textwidth]{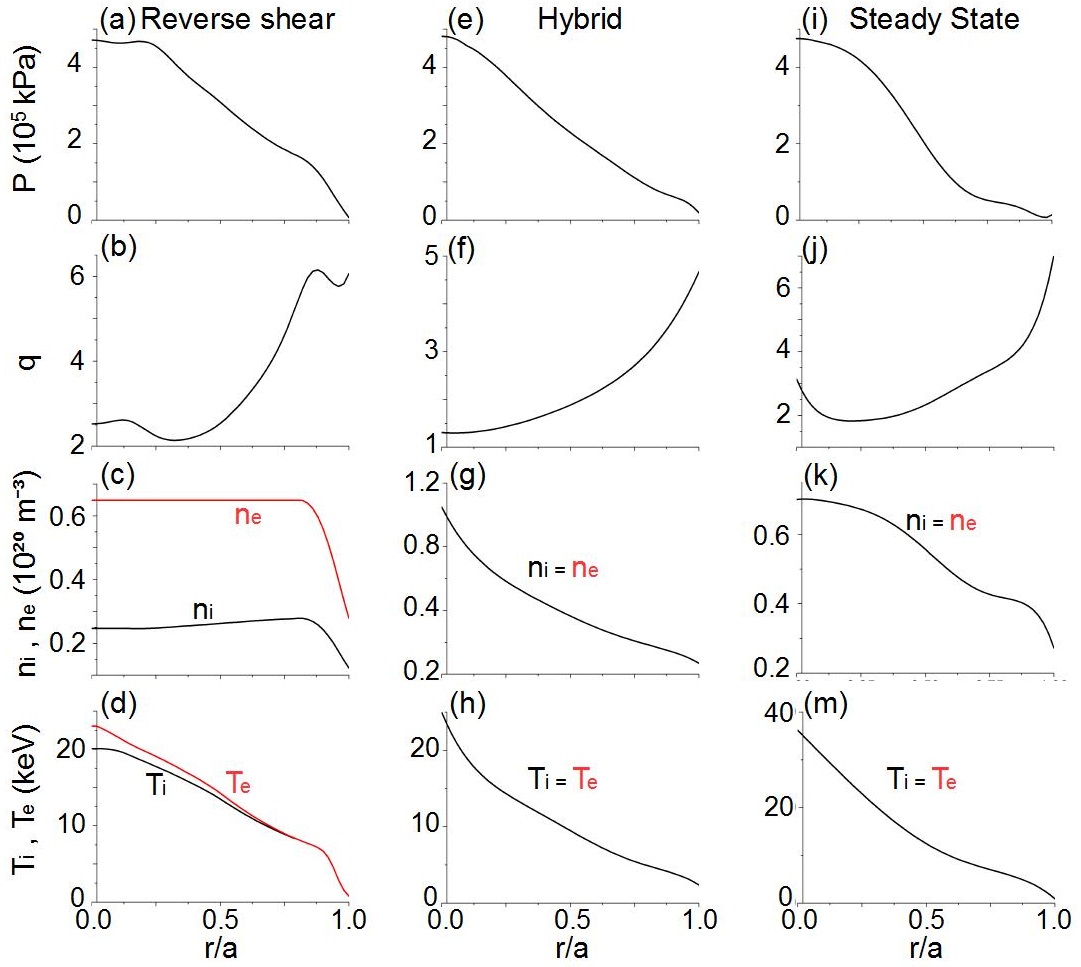}
\caption{Pressure (first row), safety factor (second row), thermal ions and electron density (third row), thermal ions and electron temperature (forth row) in the reverse shear (first column), hybrid (second column) and steady state (third column) models. }\label{FIG:1}
\end{figure*}

Figure~\ref{FIG:1} shows the models pressure, safety factor, thermal ions and electron density as well as the thermal ions and electron temperature profiles. The minimum of the safety factor profile for the reverse shear mode is located at $r/a = 0.35$ although in the SS model the minimum is located closer to the magnetic axis, at $r/a = 0.2$. The thermal electron and ion density/temperature profiles are assumed the same in the H and SS models for simplicity. The reverse shear model shows the smallest thermal ion density in the magnetic axis, $4$ times smaller regarding the H model and $3$ times compared to the SS model. It should be noted than the SS model has the highest thermal ion/electron temperature, almost a $2$ times higher regarding the reverse shear model and a $30 \%$ compared to the H model.

\begin{figure*}[h!]
\centering
\includegraphics[width=0.8\textwidth]{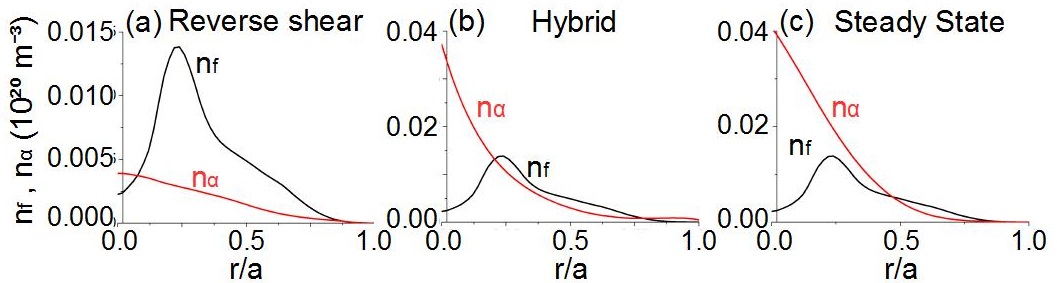}
\caption{Density profile of the NBI driven EP and alpha particles in the reverse shear (a), hybrid (b) and steady state (c) models.}\label{FIG:2}
\end{figure*}

Figure~\ref{FIG:2} shows the density profile of the NBI driven EP and alpha particles. The density profile of the NBI driven EP is the same for all the models, indicating a local maximum of the NBI EP at $r/a = 0.25$ resulting from off-axis NBI deposition. The density profile of the alpha particle component is flatter than the NBI EP and the local maximum is located at the magnetic axis. The maximum of the alpha particle density is one order of magnitude smaller with respect to the NBI EP in the RS model although it is $3$ times larger in the H and SS scenarios.

\begin{figure*}[h!]
\centering
\includegraphics[width=0.8\textwidth]{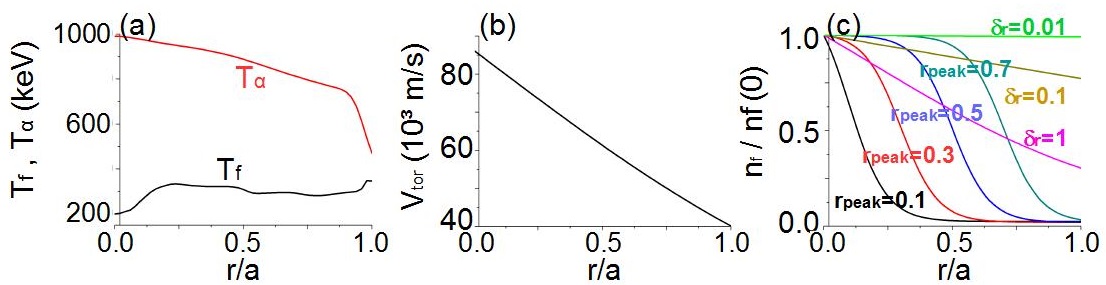}
\caption{Profiles of the NBI driven EP and alpha particle temperature in the RS model (a), equilibrium thermal plasma toroidal rotation (b) and analytic density profiles in the studies where the EP density profile is modified (c).}\label{FIG:3}
\end{figure*}

Figure~\ref{FIG:3}a shows the NBI driven EP and alpha particle temperature profiles in the RS case. The same temperature profile for the NBI driven EP is used in the H and SS models, although the alpha particle temperature is assumed flat for simplicity with $T_{\alpha}(0) = 846$ keV in the H case and $T_{\alpha}(0) = 997$ keV in the SS model. Figure~\ref{FIG:3}b shows the equilibrium toroidal rotation and figure~\ref{FIG:3}c the analytic EP density profiles used in the simulations where we study the sensitivity to variations in the NBI EP and Alpha particle density profiles on the instability growth rate and frequency. The analytic expression used is the following:

\begin{equation}
\label{EP_dens}
$$n_{f,||}(r) = \frac{(0.5 (1+ \tanh(\delta_{r} \cdot (r_{peak}-r))+0.02)}{(0.5 (1+\tanh(\delta_{r} \cdot r_{peak}))+0.02)}$$
\end{equation}
The location of the profile gradient is controlled by the parameter ($r_{peak}$) and the flatness by ($\delta_{r}$).

\begin{figure}[h!]
\centering
\includegraphics[width=0.5\textwidth]{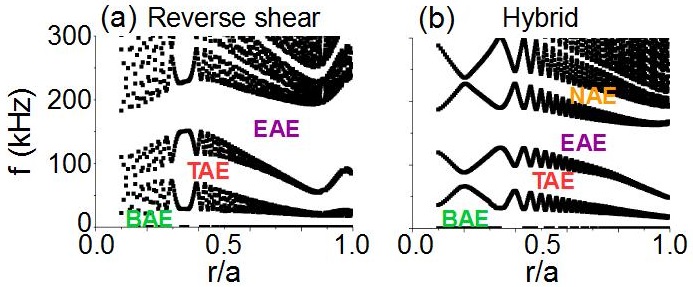}
\caption{Alfv\' en gaps of the $n=15$ toroidal mode family for the reverse shear (a) and hybrid (b) scenarios.}\label{FIG:4}
\end{figure}

Figure~\ref{FIG:4} shows the Alfv\' en gaps of $n=15$ toroidal mode family for the RS (a) and H (b) scenarios. There are three main Alfv\' en gaps: below $25$ kHz where BAE/BAAE can be destabilized, between $[40,100]$ kHz where TAE are unstable and above $100$ where the EAE can be driven. There is another gap above $200$ kHz linked to the NAE, very narrow for the RS scenario although significant for the H case. The gaps for the SS scenario are not shown because the continuum plot is similar to the RS scenario.

\subsection{Simulations parameters}

The simulations are performed with a uniform radial grid of 1000 points. The dynamic and equilibrium toroidal (n) and poloidal (m) modes included in the study are summarized in table~\ref{Table:4} for each scenario. In the following, the mode number notation is $m/n$, consistent with the $q=m/n$ definition for the associated resonance.

\begin{table}[h]
\centering
\begin{tabular}{c | c c c}
\hline
(n) & RS (m) & H (m) & SS (m)  \\ \hline
$1$ & $[2,4]$ & $[1,4]$ & $[1,4]$ \\
$2$ & $[4,8]$ & $[2,7]$ & $[3,8]$ \\
$3$ & $[6,12]$ & $[3,10]$ & $[5,12]$ \\
$4$ & $[8,16]$ & $[5,14]$ & $[7,16]$ \\
$5$ & $[10,20]$ & $[6,17]$ & $[9,20]$ \\
$6$ & $[12,24]$ & $[7,21]$ & $[10,24]$ \\ 
$7$ & $[14,28]$ & $[9,24]$ & $[12,28]$ \\ 
$8$ & $[16,32]$ & $[10,28]$ & $[14,32]$ \\ 
$9$ & $[18,36]$ & $[11,31]$ & $[16,36]$ \\ 
$10$ & $[20,40]$ & $[12,35]$ & $[18,40]$ \\ 
$11$ & $[22,44]$ & $[14,38]$ & $[19,44]$ \\ 
$12$ & $[24,48]$ & $[15,42]$ & $[21,48]$ \\ 
$13$ & $[26,52]$ & $[16,45]$ & $[23,52]$ \\ 
$14$ & $[28,56]$ & $[18,49]$ & $[25,56]$ \\ 
$15$ & $[30,60]$ & $[19,52]$ & $[27,60]$ \\ 
$16$ & $[32,64]$ & $[20,56]$ & $[28,64]$ \\ 
$17$ & $[34,68]$ & $[22,59]$ & $[30,68]$ \\ 
$18$ & $[36,72]$ & $[23,63]$ & $[32,72]$ \\ 
$19$ & $[38,76]$ & $[24,66]$ & $[34,76]$ \\ 
$20$ & $[40,80]$ & $[25,70]$ & $[36,80]$ \\  \hline
(n) & & All cases (m) &  \\ \hline
$0$ & & $[0,15]$ &  \\
\end{tabular}
\caption{Dynamic and equilibrium toroidal (n) and poloidal (m) modes.} \label{Table:4}
\end{table}

The kinetic closure moment equations (6) and (9) break the usual MHD parities. This is taken into account by including both parities $sin(m\theta + n\zeta)$ and $cos(m\theta + n\zeta)$ for all dynamic variables, and allowing for both a growth rate and real frequency in the eigenmode time series analysis. The convention of the code is, in case of the pressure eigenfunction, that $n > 0$ corresponds to the Fourier component $\cos(m\theta + n\zeta)$ and $n < 0$ to $\sin(-m\theta - n\zeta)$. For example, the Fourier component for mode $-7/2$ is $\cos(-7\theta + 2\zeta)$ and for the mode $7/-2$ is $\sin(-7\theta + 2\zeta)$. The magnetic Lundquist number is $S=5\cdot 10^6$, below the experimental value in ITER operation scenarios, so the plasma described is more resistive and the pressure gradient driven modes are artificially enhanced. A correct description of the stability of the pressure gradient driven modes requires at least a two fluid model (thermal electron and ions) to reproduce correctly the fast reconnection regime typical of a low resistivity plasma. Consequently, the analysis of the pressure gradient driven modes is out of the scope of the present study. Nevertheless, the pressure gradient driven modes calculated in the simulations are used as a reference to compare the cases with stable/unstable AEs.

The density ratio between the NBI driven energetic particles and bulk plasma ($n_{f}(0)/n_{e}(0)$) at the magnetic axis is controlled through the $\beta_{f}=$ value ($\beta_{\alpha}$ for the alpha particle), linked to the NBI injection intensity and calculated by the code TRANSP without the effect of the anomalous beam ion transport. The ratio between the NBI driven energetic particle thermal velocity and Alfv\' en velocity at the magnetic axis,$v_{th,f}/v_{A0}$ ($v_{th,\alpha}/v_{A0}$ for the alpha particles), controls the resonance efficiency between the AEs and energetic particles, associated with the EP energy. We consider a Maxwellian for the NBI driven energetic particle and alpha particles distribution functions.

\section{$\beta$ threshold of the AEs destabilized by the alpha particles \label{sec:beta}}

\begin{figure*}[h!]
\centering
\includegraphics[width=0.9\textwidth]{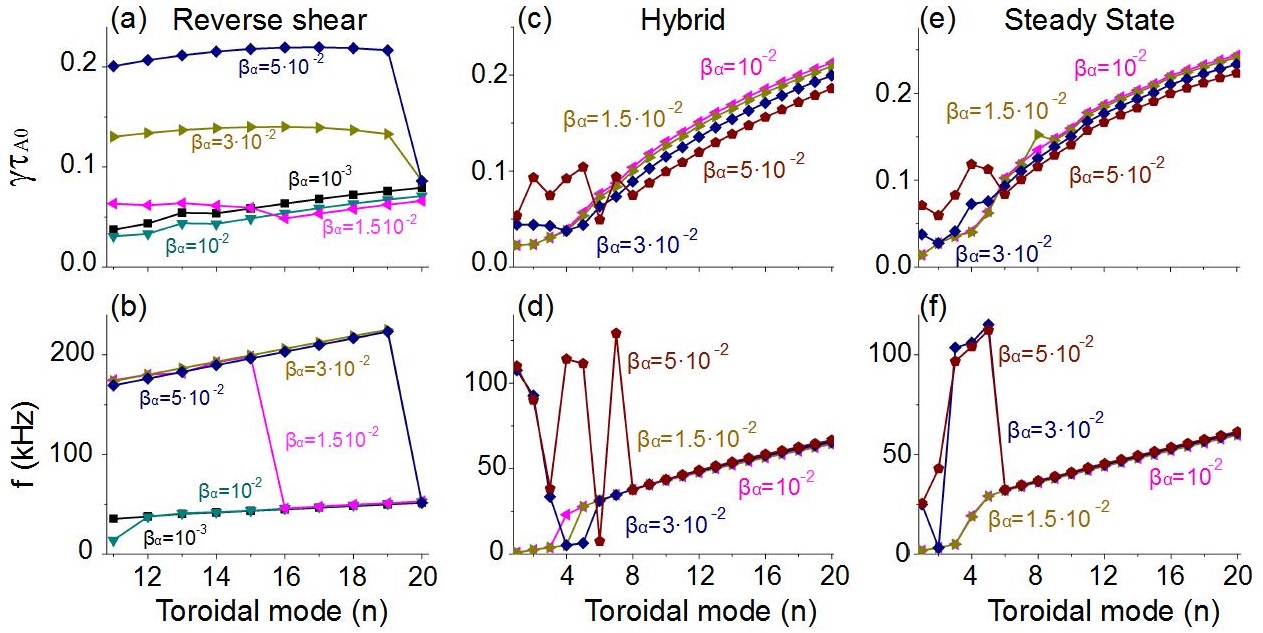}
\caption{Instability growth rate (upper row) and frequency (lower row) for different $\beta_{\alpha,0}$ values in the reverse shear (first column), hybrid (second column) and steady state (third column) model.}\label{FIG:5}
\end{figure*}

In this section we study the effect of the alpha particle $\beta_{0}$ on the instability growth rate (upper row) and frequency (bottom row) if the NBI driven EP configuration is fixed ($\beta_{f,0}=0.00073$), see figure~\ref{FIG:5}. For the RS scenario, if $\beta_{\alpha,0}$ increases up to $0.015$ ($2.5$ times above the original $\beta_{\alpha,0}$ value) the growth rate and frequency of the $n=11-15$ modes increase ranging between $f=[175,225]$ kHz. In addition, if $\beta_{\alpha,0}$ is further increased to $0.03$, the growth rate of all modes increases (except for $n=20$). In the H model, a threshold is overcome if $\beta_{\alpha,0} = 0.03$ ($3$ times the original $\beta_{\alpha,0}$ value) for the $n=1-3$ modes, up to the $n=8$ mode if $\beta_{\alpha,0} = 0.05$, with a frequency ranging between $f=[100,130]$ kHz. In the SS model, the growth rate of the $n \le 5$ modes increases if $\beta_{\alpha,0} = 0.03$ ($3$ times the original $\beta_{\alpha,0}$ value) and the frequency ranges between $f=[25,125]$ kHz. The threshold in the growth rate and frequency indicates the destabilization of an AE. The RS model shows a lower threshold in $\beta_{\alpha,0}$ because the thermal ion density is smaller (larger Alfven velocity) compared to the other models, leading to a more efficient resonance between the EP and bulk plasma to destabilize the AEs (lower $v_{th,f}/v_{A0}$ and $v_{th,\alpha}/v_{A0}$ ratios). It should be noted that the simulations with stable AEs show unstable pressure gradient driven modes, although these instabilities are most likely stable in the real operational regime of ITER plasma. These modes are destabilized in the present simulations because the magnetic Lundquist number of the model ($S=5 \cdot 10^{6}$) is below the realistic for ITER. In addition, a discussion about the consequence of adding the FLR effects in the model for simulations with unstable alpha particle driven AEs is included in the Appendix.

\begin{figure*}[h!]
\centering
\includegraphics[width=0.9\textwidth]{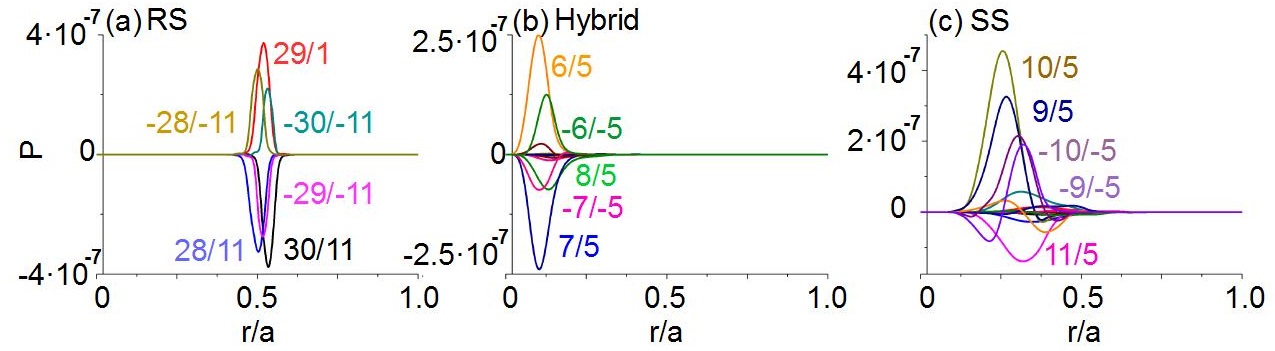}
\caption{Instability pressure eigenfunction for $\beta_{\alpha,0} = 0.05$ of $n=11$ mode in the reverse shear model (a), $n=5$ mode in the hybrid model (b) and $n=5$ mode in the steady state model.}\label{FIG:6}
\end{figure*}

Figure~\ref{FIG:6} shows the pressure eigenfunctions of instabilities above the threshold to destabilize AE caused by the alpha particles. The AE in the RS model, figure~\ref{FIG:6}a, is an $m/n=28/11 - 30/11$ EAE destabilized in the reverse magnetic shear region. The AE in the H model, figure~\ref{FIG:6}b, is a $6/5 - 7/5$ TAE destabilized near the magnetic axis. The AE in the SS model, figure~\ref{FIG:8}c, is a $10/5 - 9/5$ TAE destabilized in the inner plasma region.

\section{$\beta$ threshold of the AEs destabilized by the NBI driven EP \label{sec:beta2}}

\begin{figure*}[h!]
\centering
\includegraphics[width=0.9\textwidth]{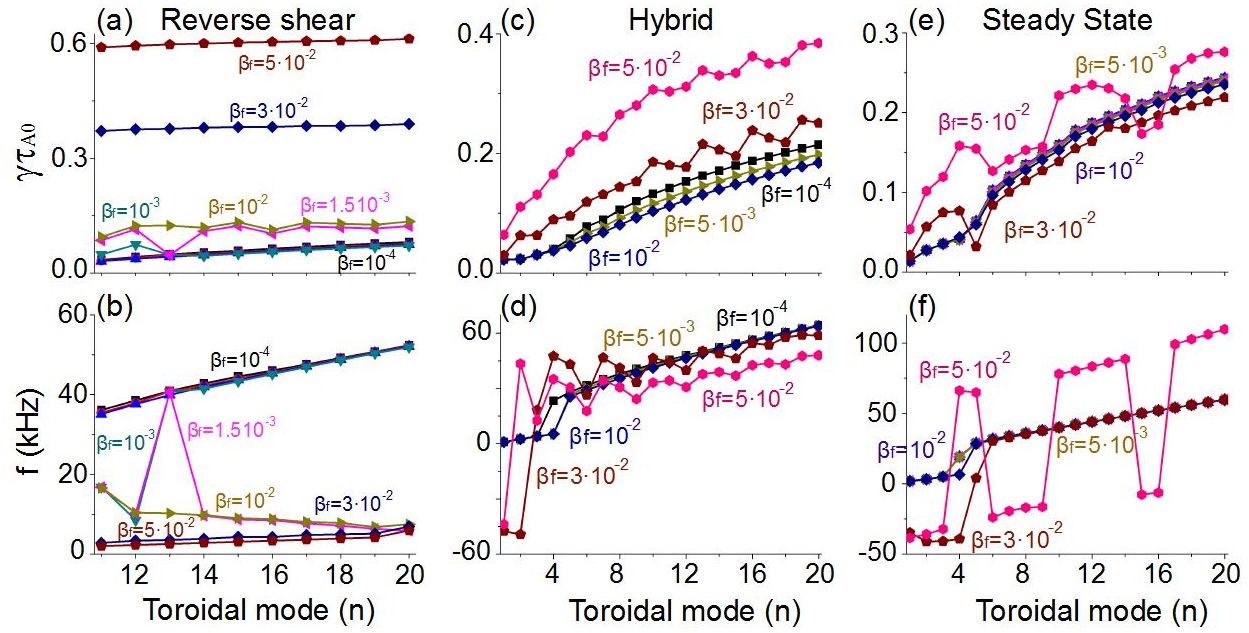}
\caption{Instability growth rate (upper row) and frequency (lower row) for different $\beta_{f,0}$ values in the reverse shear (first column), hybrid (second column) and steady state (third column) model.}\label{FIG:7}
\end{figure*}

In this section we analyze the effect of the NBI injection intensity on the instability growth rate (upper row) and frequency (bottom row) if the alpha particle configuration is fixed ($\beta_{\alpha,0}=0.0062$), see figure~\ref{FIG:7}. The threshold of the AEs in the RS model is $\beta_{f,0} = 0.0015$ ($2$ times the original $\beta_{f,0}$ value), $\beta_{f,0} = 0.03$ in the H model ($45$ times the original $\beta_{f,0}$) and $\beta_{f,0} = 0.03$ for $n<6$ (all modes if $\beta_{f,0} = 0.05$) in the SS model ($45$ times the original $\beta_{f,0}$). Again, the MHD-like modes with the lowest growth rate are observed close to the AE $\beta_{0}$ threshold and the RS scenario shows a lower $\beta_{0}$ threshold to destabilize AEs. Some of the unstable AEs show a lower frequency than the respective MHD-like modes below the threshold, because these instabilities propagate in the opposite direction from the thermal plasma toroidal rotation (Doppler shift effect). It should be noted that the symmetry properties of the model equations leads to the destabilization of AEs with the same frequency but opposite sign, required by the nature of the Alfven wave dispersion relation as well as the fact that the EP distribution function is a Maxwellian, resonating with Alfven waves in both directions. A discussion about the consequence of including the FLR effects in simulations with unstable NBI EP driven AEs is included in the Appendix.

\begin{figure*}[h!]
\centering
\includegraphics[width=0.9\textwidth]{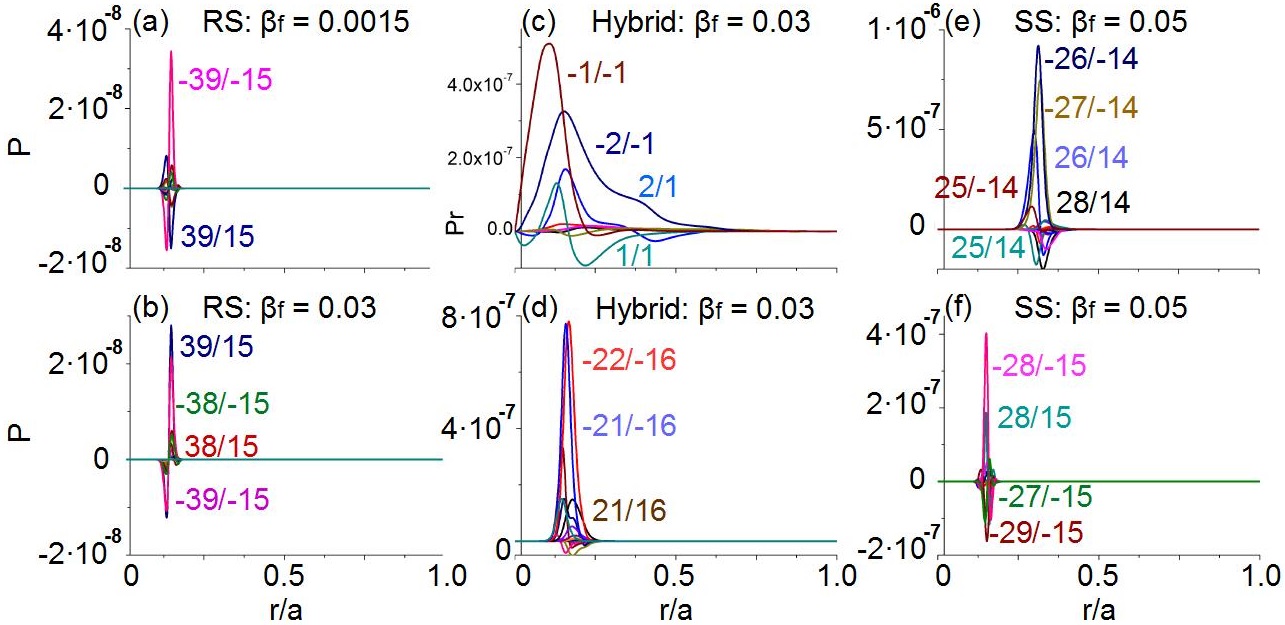}
\caption{Instability pressure eigenfunction of mode $n=15$ if $\beta_{f,0} = 0.0015$ (a) and if $\beta_{f,0} = 0.03$ (b) for the reverse shear model. Instability pressure eigenfunction if $\beta_{f,0} = 0.03$ of modes $n=1$ (c) and $n=16$ (d) for the hybrid model. Instability pressure eigenfunction if $\beta_{f,0} = 0.05$ of modes $n=14$ (c) and $n=15$ (d) for the steady state model.}\label{FIG:8}
\end{figure*}

Figure~\ref{FIG:8} shows the pressure eigenfunctions of instabilities above the threshold to destabilize AE by the NBI driven EP. In the RS model, increasing $\beta_{f,0}$ there is a transition from a $39/15$ BAE to a $39/15 - 38/15$ TAE, figure~\ref{FIG:8}a and b. In the H model, TAEs are destabilized and the low $n$ modes TAE shows a wider eigenfunction, figure~\ref{FIG:8}c and d. In the SS model, TAEs are destabilized near the magnetic axis although a $n=14$ TAE is triggered in the region where the slope of the NBI driven EP density profile is negative, while the $n=15$ TAE is triggered in the region where it is positive. This is the reason why the $n=14$ TAE propagates in the same direction as the toroidal thermal plasma rotation and the $n=15$ TAE in the opposite.

\section{Multiple EP effects on the AEs stability \label{sec:threshold}}

\begin{figure}[h!]
\centering
\includegraphics[width=0.45\textwidth]{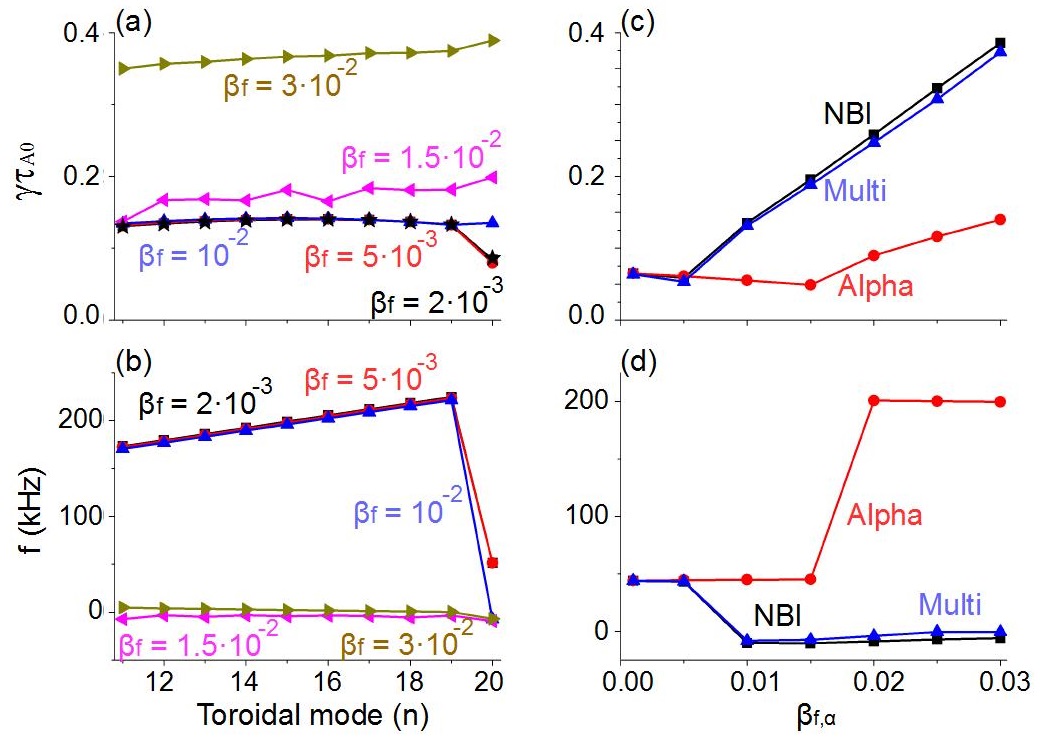}
\caption{Growth rate (a) and frequency (b) of the dominant instability for different $\beta_{f,0}$ values if the alpha particle and NBI EP driven AEs are unstable in the reverse shear model (solid lines). Growth rate (c) and frequency (d) of the dominant $n=15$ instability for different $\beta_{0}$ values of the alpha particles and NBI driven EP in simulations with only NBI driven EP (black line), only alpha particles (red line) and NBI driven EP + alpha particles (blue line).}\label{FIG:9}
\end{figure}

Now we extend the analysis to configurations where the AEs destabilized by the alpha particles and NBI EP are both unstable. To perform this study we consider only the reverse shear model because it is the scenario that shows the lower $\beta_{0}$ thresholds. The study consists of increasing the NBI injection intensity in simulations where the alpha particle AE are unstable (fixed to $\beta_{\alpha,0} = 0.03$), and then identifying the $\beta_{f,0}$ threshold that leads to dominant NBI EP driven AEs. Figure~\ref{FIG:9} panel (a) shows the growth rate and panel (b) the frequency of the dominant instability.

If $\beta_{f,0} = 0.015$ all the dominant modes evolve from EAEs to BAEs. Continuing to increase the $\beta_{f,0}$ leads to an enhancement of the BAE growth rate although the frequency is almost the same. It should be noted that the EAE growth rate is not affected by the $\beta_{f,0}$ value below the threshold, pointing out that the multiple EP effects are too weak to affect the EAE stability properties. On the other hand, if we compare the BAE growth rate in the simulations with the unstable EAE (dark yellow solid line/triangles) and stable EAE (fig~\ref{FIG:7} where $\beta_{\alpha,0}=0.0062$) for $\beta_{f,0} = 0.03$, we observed that the growth rate is $7 \%$ lower in the simulation with unstable TAE, indicating that the multiple EP effects are strong enough to affect the BAE stability properties, although not large enough to stabilize the AEs. A further analysis of the alpha particles and NBI driven EP resonances respect to the bulk plasma is performed in the Appendix, where the multiple EP effects in the AE stability are analyzed with respect to the variations in the density profile and energy of the alpha particles and NBI driven EP.

Figure~\ref{FIG:9} panel (c) shows the growth rate and panel (d) the frequency of the dominant $n=15$ instability in simulations with only NBI driven EP (black line), only alpha particles (red line) and NBI driven EP + alpha particles (blue line). In the multiple EP simulation the $\beta_{0}$ of the NBI driven EP and alpha particles is the same. The growth rate of the dominant instability in the multiple EP simulations is smaller with respect to the growth rate of the instabilities in the simulations with only NBI driven EP if $\beta_{0} > 0.01$, pointing out that the multiple EP populations damping effects are strong enough to reduce the growth rate of the dominant instability (the same trend is observed for all the modes). The same analysis is performed including the FLR effects in the model (please see the Appendix).

In summary, the stability of a plasma with multiple EP populations is different than the simple addition of the individual growth rates of the NBI driven EP and alpha particle AEs. Consequently, the study of the AE stability in reactor relevant plasma requires analysis where the multiple EP population effects are included.

\section{Conclusions and discussion \label{sec:conclusions}}

The analysis results indicate that the ITER configurations analyzed are AE stable. The threshold to destabilize AEs in the reverse shear model is $2.5$ ($2$) times higher than the $\beta_{0}$ of the alpha particle (NBI EP) regarding the reference model. In the hybrid and steady state scenarios, the $\beta_{0}$ threshold for the alpha particles (NBI EP) is $3$ ($45$) times higher than the reference model.

Above the alpha particle $\beta_{0}$ threshold, the RS model shows unstable $n=11-20$ EAE in the reverse magnetic shear region with frequencies in the range of the $f=[175,225]$ kHz. For the H scenario, $n<6$ TAE near the magnetic axis with $f=[90,130]$ kHz are destabilized. In the SS model, $n=1-2$ BAE and $n=3-5$ TAE with $f=[25,40]$ and $f=[100,115]$ kHz respectively are unstable in the inner plasma region. Above the NBI EP $\beta_{0}$ threshold, the RS model shows unstable $n=11-20$ BAE near the magnetic axis in the range of the $f=[10,20]$ kHz, as well as $n=11-19$ TAE near the magnetic axis with $f=[2,8]$ kHz if $\beta_{f,0} \ge 0.03$. The H model shows $n=2-20$ TAE near the magnetic axis with $f=[10,35]$ kHz. In the SS model $n=1-20$ TAE with $f=[60,120]$ kHz are destabilized in the inner plasma region. 

If the FLR effects are included in the RS model (see Appendix) the growth rate of the NBI EP driven AEs are reduced up to a $85 \%$ for a Larmor radius similar to the expected value for ITER ($\rho_{f} = 0.02$ m) although full stabilization is not observed. On the other hand, the alpha particle driven AEs can be stabilized for $\rho_{\alpha} = 0.05$ m (expected ITER parameter range).

The study of the AE stability in the reverse shear scenario is extended to analyze multiple EP damping effects if both alpha particle and NBI EP driven AE are unstable. In such a configuration, the growth rate of the dominant AE is lower compared to the simulations where the alpha driven AE are stable (around a $7 \%$ smaller), so the alpha particle resonance with the bulk plasma reduces the destabilizing effect of the NBI EP, although the damping effect is not strong enough to avoid triggering AEs. If the FLR effects are included (see Appendix) the multiple EP damping effects are enhanced and the growth rate of the AEs in the simulations with both NBI driven EP + alpha particles decreases up to $10 \%$.

The analysis of the resonance properties of the alpha particles and the NBI EP with the bulk plasma (see Appendix) indicates that neither of the configurations tested lead to the multiple EP interaction regime, because the AE are not stabilized and the AE growth rate in the multiple EP simulations is above the AE growth rate of the single alpha particle or single NBI EP simulations. Nevertheless, optimization trends are identified to reduce the AEs growth rate in the multiple EP configurations. In particular, we specify the conditions for the full stabilization of the AEs with respect to the density profiles of the alpha particle and NBI EP.

In summary, if the NBI injection intensity and the alpha particle production rate are large enough to overcome the AEs stability threshold, the multiple EP damping effects reduce the growth rate of the dominant AE, pointing out another optimization trend linked to the difference resonance of the EP populations with the bulk plasma. Nevertheless, the AEs destabilized individually by the alpha particles and NBI EP can be stabilized if the density gradient is located at the plasma periphery ($r/a > 0.5$), although that is not likely in ITER plasma, particularly for the alpha particles mainly generated in the plasma core. On the other hand, alpha particle and NBI EP driven AEs are also stable if the density profile gradient is weak enough (below $\delta_{r} = 0.5$). 

It should be noted that the multiple EP damping effects can be analyzed in present tokamak and stellarators during discharges where the plasma is heated by multiple NBIs with different configurations. Analysis of such configurations should lead to a better understanding of the AE stability pending future ITER operations with multiple EP populations.One example is DIII-D device, where the multiple EP population effects caused by multiple NBI injection can be studied thanks to the possibility of modifying the NBI voltage (related to EP energy) and tilt (modification of the EP density profile) \cite{96}.

\section*{Appendix}

\subsection*{Study of the alpha particle/NBI driven EP resonance with the bulk plasma}

The resonance of the alpha particles and NBI driven EP with the bulk plasma and the effect on the AEs growth rate and frequency in the RS scenario is further analyzed in this section. The aim of the study is to search for the ITER operation regime that shows a lower AE growth rate.

\begin{figure*}[h!]
\centering
\includegraphics[width=0.9\textwidth]{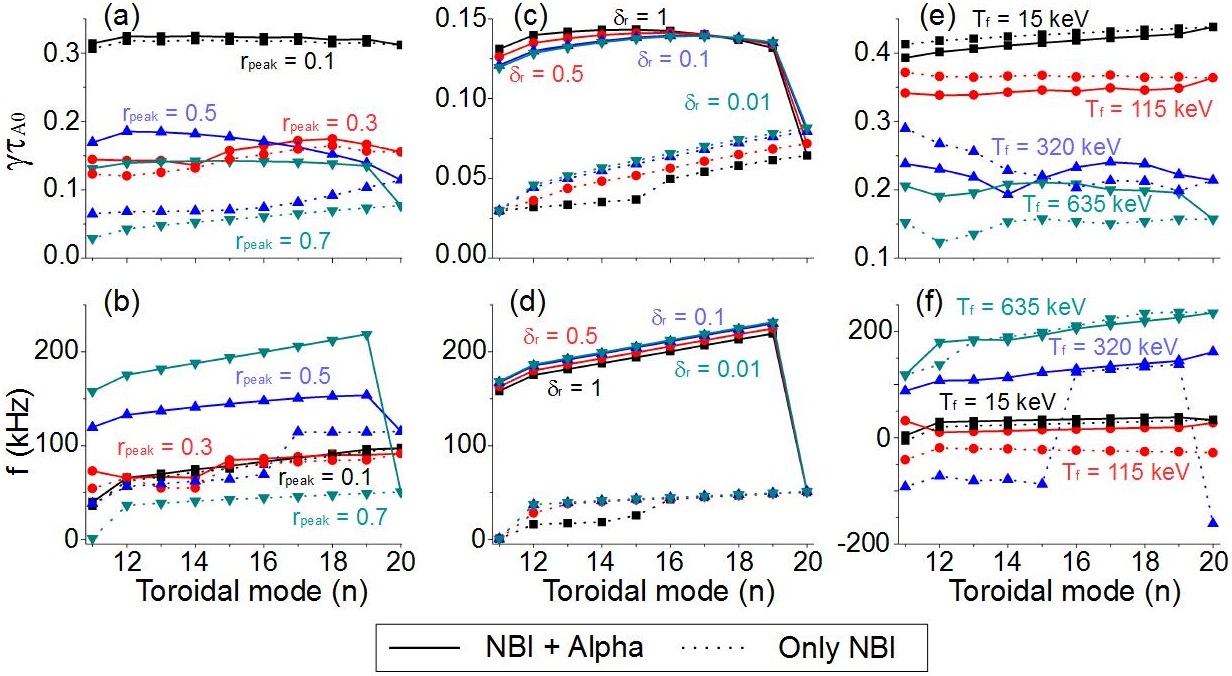}
\caption{Growth rate (upper row) and frequency (bottom row) for different NBI EP $r_{peak}$ (first column), $\delta_{r}$ (second column) and energy (third column) values. The solid lines show the multiple EP simulations and the dotted lines the simulations with only NBI driven EP.}\label{FIG:10}
\end{figure*}

First, the resonance properties of the NBI driven EP are analyzed. Figure~\ref{FIG:10} shows the AE growth rate and frequency for different density profiles and energies. It should be noted that the NBI energy in ITER is $1$ MeV, although we consider other energies as a theoretical exercise. The AE growth rates in the multiple EP simulations are higher compared to the simulations with only alpha particles and only NBI EP (see fig~\ref{FIG:5} and~\ref{FIG:7}), so none of the configurations tested lead to the stabilization of the AEs, although the instability growth rate can be reduced if the NBI is deposited off-axis, reaching a local minimum for $r_{peak} = 0.3$, in panels a and b. In that case the multiple EP simulations show almost the same AE growth rate as the simulations with only alpha particles. Modifying the flatness of the NBI EP density profile leads to the stabilization of the AEs caused by the NBI EP for any $\delta_{r}$ value, panels c and d, so only the AEs driven by the alpha particles are observed. On the other hand, if the NBI energy is modified, panels e and f, the multiple EP simulations show a lower growth rate compared to the single NBI simulations if $T_{f} = 15$ and $115$ keV as well as if $320$ keV for $n < 15$ modes. Only in the cases with $T_{f} = 115$ and $320$ keV is the AEs growth rate lower with respect to the simulations with stable alpha driven AEs.

\begin{figure*}[h!]
\centering
\includegraphics[width=0.9\textwidth]{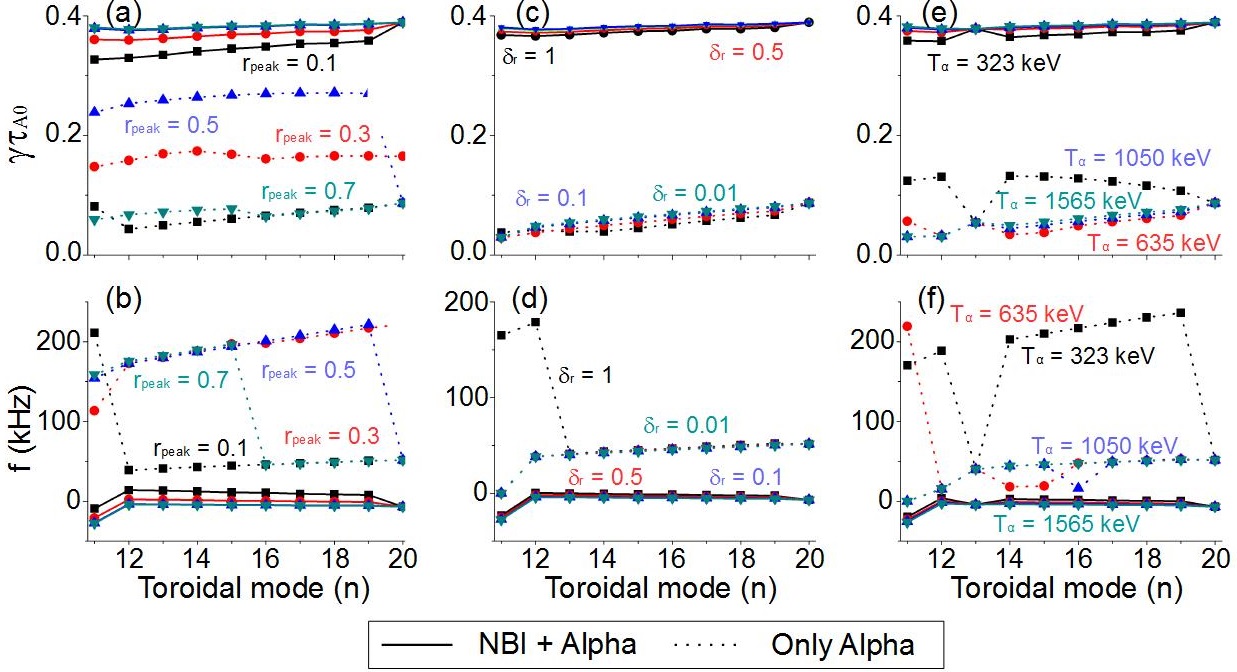}
\caption{Growth rate (upper row) and frequency (bottom row) for different alpha particle $r_{peak}$ (first column), $\delta_{r}$ (second column) and energy (third column) values. The solid lines show the multiple EP simulations and the dotted lines the simulations with only alpha particle driven EP.}\label{FIG:11}
\end{figure*}

The same analysis is made for the resonance properties of the alpha particles. As in the previous study, the growth rate of the AEs in the multiple EP simulations is above the simulations with only alpha particles (see fig~\ref{FIG:5}). On the other hand, if the alpha particles are located near the magnetic axis ($r_{peak} = 0.1$), the AEs growth rate decreases around a $25 \%$ compared to the simulations with stable alpha driven AEs. The effect of modifying the profile flatness or the alpha particle temperature is negligible, showing only an slightly decrease of the AEs growth rate for the lowest energy tested ($323$ keV).

In summary, stabilization of the AEs is not observed in any of the configurations tested. On the other hand, several optimization trends are identified leading to the reduction of the AEs growth rate regarding the simulation with stable alpha driven AEs. It should be noted that the AEs driven by the NBI EP can be stabilized if the gradient of the density profiles is located in the plasma periphery ($r_{peak} = 0.7$) or the profile gradient is reduced ($\delta_{r} \le 0.5$). Likewise, the AEs driven by the alpha particles are stabilized if the profile gradient is reduced ($\delta_{r} \le 1$).

\subsection*{Finite Larmor radius effects on the AE stability}

\begin{figure*}[h!]
\centering
\includegraphics[width=0.9\textwidth]{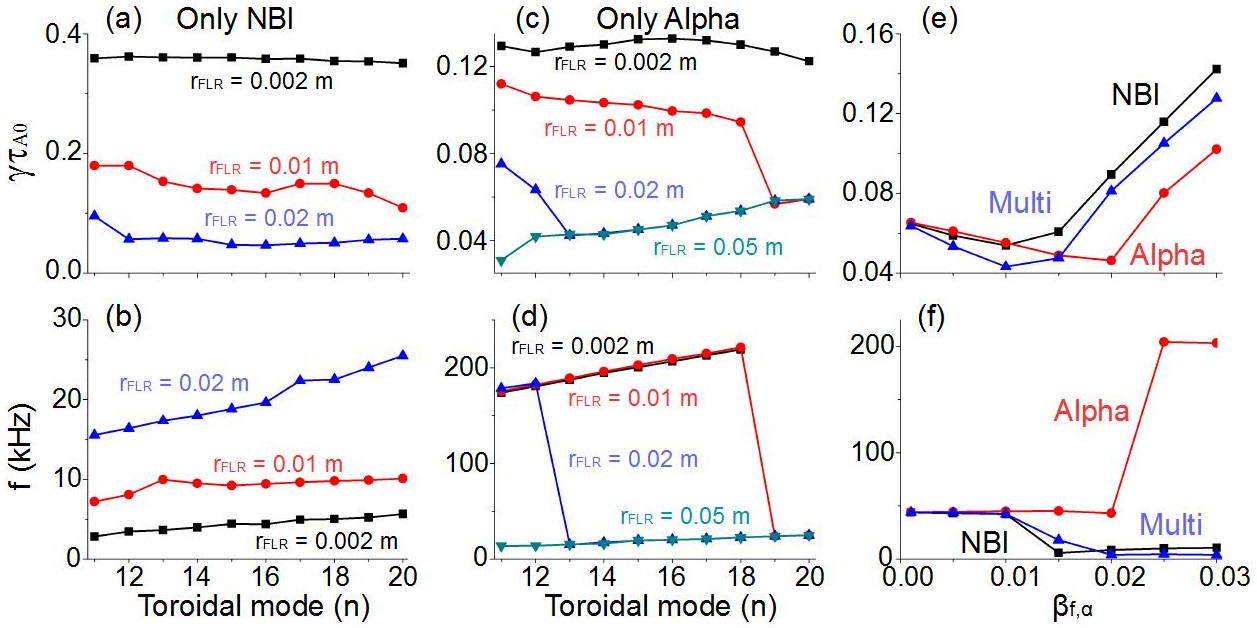}
\caption{Growth rate (a) and frequency (b) of the instability in simulations with only NBI driven EP for $\beta_{f,0} = 0.03$ with different FLR radius. Growth rate (c) and frequency (d) of the instability in simulations with only alpha particles for $\beta_{\alpha,0} = 0.03$ with different FLR radius. Growth rate (e) and frequency (f) of the dominant $n=15$ instability for different $\beta_{0}$ values of the alpha particles and NBI driven EP in simulations with only NBI driven EP (black line), only alpha particles (red line) and NBI driven EP + alpha particles (blue line). The analysis is performed for the RS operational scenario of ITER.}\label{FIG:12}
\end{figure*}

Introducing the finite Larmor radius effects (FLR) of ions and energetic particles in the numerical model leads to a reduced drive for the instabilities, reducing the growth rate or even stabilizing the AEs and pressure gradient driven modes observed in the simulations. The FLR effects are included in the equations describing the perturbation evolution in thermal plasma, NBI driven EP and alpha particles. Figure~\ref{FIG:12} shows the growth rate and frequency of the AEs destabilized by the NBI driven AE ($\beta_{f,0}=0.03$) and alpha particles ($\beta_{\alpha,0}=0.03$) if the FLR effects are included for the RS case. The expected Larmor radius in ITER operation scenarios for the alpha particles is around $0.05$ m, $0.02$ m for the NBI driven EP and $0.002$ m for the thermal ions. The simulations show that the FLR effects can reduce the growth rate of the AEs destabilized by the NBI driven EP (panels a and b), up to a $85 \%$ if $\rho_{f} = 0.02$ m, although complete stabilization is not observed. On the other hand, the AEs driven by the alpha particles (panels c and d) are stabilized for all the modes if $\rho_{\alpha} = 0.05$ m although only the $n=11$ and $12$ EAE are unstable if $\rho_{\alpha} = 0.02$ m and all the modes are AE unstable except $n=19$ and $20$ if $\rho_{\alpha} = 0.01$ m.

Figure~\ref{FIG:12}, panels e and f, shows an analysis of the multiple EP damping effects if the FLR effects are included ($\rho_{\alpha} = \rho_{f} = 0.01$ m) in the RS operational scenario of ITER. The Larmor radius used in the analysis are slightly below the expected values for NBI driven EP and alpha particles in ITER operational regime, because we want to avoid the stabilization of the AEs destabilized by each EP species due to the FLR effects. The growth rate of the AEs destabilized in the simulations ($\beta_{f,\alpha,0} \geq 0.01$) with only NBI driven EP (black line) and in the simulations with both  NBI driven EP + alpha particles (blue line) is half compared with the simulations without FLR effects. Again, the multiple EP damping effects lead to a reduction of the AEs growth rate respect to the simulations with only NBI driven EP (interaction regime), around a $10 \%$ lower, a large decrease compared with the simulation without FLR. Consequently, the FLR effects can enhance the mutiple EP damping effects in operational regimes where both NBI driven EP and alpha particles destabilize AEs.

\ack
This material based on work is partially supported both by the U.S. Department of Energy, Office of Science, under Contract DE-AC05-00OR22725 with UT-Battelle, LLC and U.S. Department of Energy, Oﬃce of Science, Oﬃce of Fusion Energy Sciences, using the DIII-D National Fusion Facility, a DOE Oﬃce of Science user facility, under Award No. DE-FC02-04ER54698. DIII-D data shown in this paper can be obtained in digital format by following the links at https://fusion.gat.com/global/D3D\_DMP. This research was sponsored in part by the Ministerio of Economia y Competitividad of Spain under project no. ENE2015-68265-P. The authors would like to thanks Y. Todo for fruitful discussions

\hfill \break

\end{document}